\documentclass{article}
\usepackage[a4paper, total={6.5in,8.6in}]{geometry}
\usepackage{setspace} 
\usepackage{mathrsfs} 
\usepackage{bbm}
\usepackage{dutchcal}
\usepackage{verbatim}
\usepackage{amsmath}
\usepackage{amsthm}
\usepackage{amssymb}
\usepackage{mathtools}
\usepackage{bm}
\usepackage{mathabx}
\usepackage{multirow}
\usepackage{graphicx}
\usepackage{subfigure}
\usepackage{float}
\usepackage{comment}
\usepackage{amsfonts}
\usepackage{color}
\usepackage[dvipsnames]{xcolor}
\usepackage{xr-hyper}
\usepackage[colorlinks=true,linkcolor=blue,citecolor=blue,pdfborder={0 0 0}]{hyperref}
\usepackage[backend=biber,style=authoryear,citestyle=authoryear,maxbibnames=99,maxcitenames=2, mincitenames=1,natbib=true,citetracker=true,useprefix=true]{biblatex}
\usepackage[affil-it]{authblk}

\graphicspath{{Images20201227/}}

\numberwithin{equation}{section}

\theoremstyle{plain}
\newtheorem{Assump}{Assumption}[section]

\newtheorem{Theo}{Theorem}[section]
\newtheorem{Prop}{Proposition}[section]
\newtheorem{Lemma}{Lemma}[section]
\newtheorem{Coro}{Corollary}[section]

\newtheorem{Def}{Definition}[section]

\theoremstyle{remark}

\newtheorem{Remark}{Remark}[section]

\newcommand{\changeN}[1]{{\color{black}{#1}}}

\newcommand{\op}[1]{o_{\Prob}({#1})}
\newcommand{\Op}[1]{O_{\Prob}({#1})}
\DeclareMathOperator{\Bias}{Bias}
\DeclareMathOperator{\Var}{Var}

\newcommand{\talpha}{\widetilde{\alpha}}

\newcommand{\hb}{\hat{b}}
\newcommand{\tb}{\tilde{b}}
\newcommand{\hDelta}{\widehat{\Delta}}
\newcommand{\tDelta}{\widetilde{\Delta}}
\newcommand{\be}{\bm{e}}
\newcommand{\ep}{\epsilon}
\newcommand{\cK}{\mathcal{K}}
\newcommand{\cL}{\mathcal{L}}
\newcommand{\hL}{\widehat{L}}

\newcommand{\mL}{\mathring{L}}
\newcommand{\tL}{\widetilde{L}}
\newcommand{\hcL}{\widehat{\mathcal{L}}}
\newcommand{\tcL}{\widetilde{\mathcal{L}}}
\newcommand{\ccL}{\check{\mathcal{L}}}

\newcommand{\bbN}{\mathbbm{N}}
\newcommand{\mrho}{\mathring{\rho}}
\newcommand{\hrho}{\hat{\rho}}
\newcommand{\trho}{\widetilde{\rho}}
\newcommand{\bbR}{\mathbbm{R}}
\newcommand{\hRSS}{\widehat{RSS}}
\newcommand{\tRSS}{\widetilde{RSS}}
\newcommand{\ctheta}{\check{\theta}}
\newcommand{\bu}{\bm{u}}
\newcommand{\bx}{\bm{x}}
\newcommand{\bX}{\bm{X}}
\newcommand{\cX}{\mathcal{X}}
\newcommand{\by}{\bm{y}}
\newcommand{\bbZ}{\mathbbm{Z}}
\newcommand{\tZ}{\tilde{Z}}

\newcommand{\ind}{\mathbbm{1}}
\newcommand*\diff{\mathop{}\!\mathrm{d}}
\newcommand{\Exp}{\mathbb{E}}
\newcommand{\Prob}{\mathbb{P}}
\newcommand{\argmin}{\operatornamewithlimits{\arg\min}}
\newcommand{\argmax}{\operatornamewithlimits{\arg\max}}

\newcommand{\floor}[1]{\lfloor{#1}\rfloor}

\newcommand{\mathph}[1]{\mathrel{\phantom{#1}}}
\newcommand{\pto}{\xrightarrow{p}}

\title{Estimating POT Second-order Parameter for Bias Correction}
\date{}
\author{Nan Zou\thanks{Email: nan.zou@mq.edu.au}}
\affil{School of Mathematical and Physical Sciences,
Macquarie University, Macquarie Park, NSW, Australia}

\begin{document}
\maketitle

\abstract{
The stable tail dependence function provides a full characterization of the extremal dependence structures. Unfortunately, the estimation of the stable tail dependence function often suffers from significant bias, whose scale relates to the Peaks-Over-Threshold (POT) second-order parameter. For this second-order parameter, this paper introduces a penalized estimator that discourages it from being too close to zero. This paper then establishes this estimator's asymptotic consistency, uses it to correct the bias in the estimation of the stable tail dependence function, and illustrates its desirable empirical properties in the estimation of the extremal dependence structures.  
}


\section{Introduction}


Misfortunes never come singly, and extreme events come together frequently. To depict the dependence structure of extreme events, extreme value theory provides many characterizations, such as the spectral measure (\textcite{de1977limit}), the spectral distribution function (\textcite{einmahl1997estimating}), the extreme-value copula (\textcite{gudendorf2010extreme}), and the Pickands dependence function (\textcite{Pic81}). In this paper, we focus on the stable tail dependence function (\textcite{huang1992statistics}) defined as follows. Suppose that $(\bX_{i})_{i\in \bbZ}$ is a sequence of i.i.d. $d$-dimensional random vectors, in which $(\bX_{i})_{i=1,\dots,n}$ is observable. Suppose that $\bX_{i} = (X_{i}^{(1)},\dots,X_{i}^{(d)})'$, $i=1,\dots,n$. Let $\bx = (x_{1},\dots x_{d})'$. Let
$F(\bx)=\Prob(X_{1}^{(1)}\leq x_{1},\dots, X_{1}^{(d)}\leq x_{d})$ and $F_{j}(x)=\Prob(X_{1}^{(j)}\leq x)$, $j=1,\dots,d$, be the joint and marginal cumulative distribution functions of $\bX_{1}$. Assume that $F_{j}$ is continuous for $j=1,\dots,d$ and let $F_{j}^{\leftarrow}(t)=\inf\{z\in \bbR: F_{j}(z)\geq t\}$. Let the stable tail dependence function be the function $L$ that satisfies Assumption \ref{assump:firorder} below.
\begin{Assump}[First-order condition]\label{assump:firorder}
Assume that there exists a non-degenerate function $L$ on $[0,\infty)^{d}$, such that for all $T>0$,
\[
\lim_{t\to \infty}\sup_{\bx\in[0,T]^{d}}\bigg|t\Big[1-F\big(F_{1}^{\leftarrow}(1-t^{-1}x_{1}),\dots,F_{d}^{\leftarrow}(1-t^{-1}x_{d})\big)\Big]-L(\bx)\bigg|=0.
\]
\end{Assump}
Assumption \ref{assump:firorder} in a sense equates to the assumption that $F$ is in the domain of attraction of an extreme value distribution (\textcite[p. 903]{fougeres2015bias}). Assumption \ref{assump:firorder} also indicates that $L$ is a characterization of the dependence structure of extreme events. To estimate $L$, 
for each $j=1,\dots,d$, let $X_{1,n}^{(j)}\leq X_{2,n}^{(j)} \leq \dots \leq X_{n,n}^{(j)}$ be the order statistics of $\{X_{i}^{(j)}\}_{i=1,\dots,n}$, let $k=k_{n}$ be a sequence of positive real numbers such that $k\to \infty$ and $k/n\to 0$ as $n\to \infty$, and let $\floor{\xi}$ be the largest integer smaller or equal to $\xi$. Let the 
(non-bias-corrected)
estimator of the stable tail dependence function $L$ be defined by, for $\bx \in [0,\infty)^{d}$,
\begin{equation}\label{eq:hatL}
\hL_{k}(\bx)=\frac{1}{k}\sum_{i=1}^{n}\ind_{\big\{X_{i}^{(1)}\geq X_{n-\floor{kx_{1}}+1,n}^{(1)} \ \text{or}\ \dots \ \text{or} \  X_{i}^{(d)}\geq X_{n-\floor{kx_{d}}+1,n}^{(d)}\big\}}.
\end{equation}
By e.g., \textcite[p. 904]{fougeres2015bias}, when the intermediate sequence $k=k_{n}$ in \eqref{eq:hatL} is relatively large, $\hL_{k}$'s asymptotic bias dominates its asymptotic standard deviation.
Indeed, by e.g., Proposition \ref{prop:L_k} or \textcite[p. 904]{fougeres2015bias}, under Assumption \ref{assump:secorder} below and additional conditions, the asymptotic bias of $\hL_{k}$ has an order of $\alpha(n/k)$; more specifically, with $\alpha$ and $M$ given in Assumption \ref{assump:secorder},
\begin{equation}\label{eq:biasHatL}
   \Bias(\hL_{k}(\bx)) = \alpha(n/k)M(\bx) + o(\alpha(n/k)).
\end{equation}
\begin{Assump}[Second-order condition]\label{assump:secorder}
Assume that there exists a positive function $\alpha$ and a non-null function $M$ such that $\lim_{t\to \infty}\alpha(t)=0$ and for all $T>0$, 
\[
\lim_{t\to \infty}\sup_{\bx\in[0,T]^{d}}\Bigg|\frac{t\Big[1-F\big(F_{1}^{\leftarrow}(1-t^{-1}x_{1}),\dots,F_{d}^{\leftarrow}(1-t^{-1}x_{d})\big)\Big]-L(\bx)}{\alpha(t)}-M(\bx)\Bigg|=0.
\]
\end{Assump}

\noindent
Given Assumption \ref{assump:secorder}, by e.g., Remarks 1 and 3 of \textcite{fougeres2015bias} and Lemma 2.2 of \textcite{bucher2019second}, there exists some $\rho\leq 0$ such that for all $a>0$, $\bx\in [0,\infty)^{d}$, we have 
\begin{align}\label{eq:regv}
    L(a\bx)&=aL(\bx), & \lim_{m\to \infty}\alpha(ma)/\alpha(m)&=a^{\rho}, & M(a\bx)&=a^{1-\rho}M(\bx). 
\end{align}
To bring down $\hL_{k}$'s asymptotic bias in \eqref{eq:biasHatL}, many bias-corrected estimators, e.g., \textcite{peng2010practical},
\textcite{fougeres2015bias}, \textcite{beirlant2016bias}, \textcite{goegebeur2017kernel}, \textcite{escobar2017bias}, \textcite{goegebeur2022robust}, have leveraged the regular varying and homogeneous properties in \eqref{eq:regv}. For example, when the true second-order parameter $\rho$ is known, \textcite[p. 909]{fougeres2015bias}'s dot estimator of stable tail dependence function $L$ is defined by
\begin{equation}\label{eq:Fou_dot_L}
\mL_{k,a}(\bx,\rho)=\hL_{ka}(\bx)-\hL_{k(a^{-\rho}+1)^{-1/\rho}}(\bx)+\hL_{k}(\bx).
\end{equation}
Under some conditions, by \eqref{eq:biasHatL} and \eqref{eq:regv}, compared to \eqref{eq:biasHatL}, \textcite{fougeres2015bias}'s dot estimator in \eqref{eq:Fou_dot_L} has a smaller order of asymptotic bias; specifically,
\begin{align*}
\Bias(\mL_{k,a}(\bx,\rho))&=\Bias(\hL_{ka}(\bx))-\Bias (\hL_{k (a^{-\rho}+1)^{-1/\rho}}(\bx))+\Bias(\hL_{k}(\bx))\\
    &= \left[\alpha(n/(ka))-\alpha(n/(k (a^{-\rho}+1)^{-1/\rho}))+ \alpha(n/k)\right] M(\bx)  + o(\alpha(n/k))
    \\
    &=\left[a^{-\rho}-(a^{-\rho}+1)+1\right]\alpha(n/k)M(\bx) + o(\alpha(n/k)) =  o(\alpha(n/k)). 
\end{align*}
However, in practice, the second-order parameter $\rho$ in \eqref{eq:regv} is usually unknown. To estimate this second-order parameter $\rho$, \textcite{fougeres2015bias}, \textcite{beirlant2016bias}, and \textcite{goegebeur2017kernel} designed estimators that are consistent under certain conditions. Nevertheless, empirically, these estimators of $\rho$ may get too close to zero from time to time; see \textcite[p. 926]{fougeres2015bias} and \textcite[p. 457]{beirlant2016bias}. When the estimator of $\rho$ gets too close to zero, the ensuing bias-corrected estimator of $L$ may be very far from the true $L$. Let us take \textcite{fougeres2015bias} dot estimator $\mL_{k,a}(\bx,\gamma)$ defined in \eqref{eq:Fou_dot_L} as an example again. Let $a\in (0,1)$ as in \textcite[p. 914]{fougeres2015bias}. Suppose that $\rho<0$ is the true second-order parameter and $\gamma<0$ is an approximation of $\rho$. Under some conditions, by \eqref{eq:biasHatL} and \eqref{eq:regv}, 
    \[
    \Bias(\mL_{k,a}(\bx,\gamma))
    =\left[a^{-\rho}-(a^{-\gamma}+1)^{|\rho|/|\gamma|}+1\right]\alpha(n/k)M(\bx) + o(\alpha(n/k))\to 
    \begin{cases}
         \infty \ \text{or} \ -\infty,  & \gamma\uparrow 0 \\
            O(\alpha(n/k)), & \gamma\to -\infty
    \end{cases}
    \]


In a word, after we replace the true second-order parameter $\rho$ by its approximate $\gamma$, the asymptotic absolute bias of \textcite{fougeres2015bias} dot estimator $\mL_{k,a}(\bx,\gamma)$ will at least have the same order as the non-bias-corrected estimator when $\gamma\to -\infty$ but will blow up when $\gamma\uparrow 0$. This kind of asymmetric behavior in
the performance of the bias-corrected estimators suggests that we could improve the performance of the bias-corrected estimators by driving $\gamma$ away from zero. 

For this sake, we first develop a penalized, nonlinear least square estimator of the true second-order parameter $\rho$, where we intend to use the penalty to discourage the estimator to get too close to zero. We then establish the consistency of this penalized estimator with a functional central limit theorem for $\hL_{ka}(\bx)$ that is uniform not only in $\bx$ but also in $a$. Finally, we plug this penalized estimator into a collection of bias-corrected estimators of $L$ and briefly analyze the theoretical and empirical performance of these bias-corrected estimators. 

The remaining part of this paper is organized as follows. In Section \ref{sect:pen}, we give further details in the motivation and the definition of the penalized estimator of $\rho$ and then discuss its asymptotic properties. In Section \ref{sect:estL}, we review some bias-corrected estimators of $L$, which all depend on the estimator of $\rho$. In Section \ref{sect:simul}, we plug the penalized estimator of $\rho$ in  Section \ref{sect:pen} into the bias-corrected estimators of $L$ in \ref{sect:estL} and illustrate the performance of these plug-in estimators. Appendix includes all the proofs.

\section{Methodology}



\subsection{Preliminaries}
To further motivate the penalized estimator of the second-order parameter $\rho$, let us first formally review the asymptotic behaviors of $\hL_{k}$ defined in \eqref{eq:hatL} under some additional assumptions below.

\begin{Assump}[Third-order condition]\label{assump:thiorder}
Assume that there exist a positive function $\beta$ and a non-null function $N$ such that $\lim_{t\to \infty}\beta(t)=0$, $N$ is not a multiple of $M$, and for all $T>0$, 
\[
\lim_{t\to \infty}\sup_{\bx\in[0,T]^{d}}\left|\frac{1}{\beta(t)}
\Bigg\{\frac{t\Big[1-F\big(F_{1}^{\leftarrow}(1-t^{-1}x_{1}),\dots,F_{d}^{\leftarrow}(1-t^{-1}x_{d})\big)\Big]-L(\bx)}{\alpha(t)}-M(\bx)\Bigg\}-N(\bx)\right|=0.
\]
\end{Assump}

\begin{Remark}
In addition to $L$, $\alpha$, and $M$ in \eqref{eq:regv}, $\beta$ and $N$ also have regular varying or homogeneous properties under Assumption \ref{assump:thiorder}. Specifically, by \textcite[Remark 3]{fougeres2015bias}, there exists some $\rho'\leq 0$ such that for all $a>0$, $\bx\in [0,\infty)^{d}$, we have
\begin{equation}\label{eq:regv2}
\begin{aligned}
    \lim_{m\to \infty}\beta(ma)/\beta(m)&=a^{\rho'}, & N(a\bx)&=a^{1-\rho-\rho'}N(\bx). 
\end{aligned}
\end{equation}
\end{Remark}

\begin{Assump}[Strict Negativity of $\rho$ and $\rho'$]
\label{assump:negativity}
Assume that $\rho$ in \eqref{eq:regv} and $\rho'$ in \eqref{eq:regv2} satisfy $\rho<0$ and $\rho'<0$.
\end{Assump}

\begin{Assump}[Smoothness of $L$, $M$, and $N$]\label{assump:smoothness}
Assume that the partial derivatives of $L$ exist and denote the partial derivative with respect to the $j$-th coordinate by $\partial_{j}L$. Assume that for $j=1,\dots,d$, $\partial_{j}L$ is continuous on $\{\bx\in[0,\infty)^{d}:x_{j}>0\}$. Further, assume that the function $M$ in Assumption \ref{assump:secorder} is differentiable and the function $N$ in Assumption \ref{assump:thiorder} is continuous.
\end{Assump}

\begin{Assump}[Speed of the intermediate sequence $k$]\label{assump:kspeed} Assume 
\begin{itemize}
    \item $k\to \infty$, $k/n\to 0$.
    \item $\sqrt{k}\alpha(n/k)\to \infty$, $\sqrt{k}\alpha(n/k)\beta(n/k)\to 0$.
\end{itemize}
\end{Assump}
\begin{Remark}
Indeed, the order of the asymptotic standard deviation of $\hL_{k}$ and the bias of $\hL_{k}$ is $1/\sqrt{k}$ and $\alpha(n/k)$, respectively; see e.g., \textcite[Proposition 2]{fougeres2015bias} and Proposition \ref{prop:L_k}. Hence, Assumption \ref{assump:kspeed} assumes asymptotically the bias of $\hL_{k}$ dominates its standard deviation. 
\end{Remark}
Now let us describe the asymptotic distribution of the rescaled version of $\hL_{k}$. Let $\be_{j}$ be a $d$-dimensional vector with value 1 at its $j$-th coordinate and zeros elsewhere. Define
\begin{equation}\label{eq:ZL}
Z_{L}(\bx)=W_{L}(\bx)-\sum_{j=1}^{d}W_{L}(x_{j}\be_{j})\partial_{j} L(\bx),
\end{equation}
where $W_{L}$ is a Gaussian process on $[0, \infty)^{d}$ with continuous paths, zero mean, and covariance structure 
\[
\Exp[W_{L}(\bx)W_{L}(\by)]=\mu \{R(\bx)\cap R(\by)\},
\]
where 
\[
R(\bx)= \{\bu\in [0,\infty)^{d}: 0\leq u_{j}\leq x_{j} \ \text{for some} \ j=1,\dots,d\},
\]
and $\mu$ is the measure defined by $\mu\{A(\bx)\}=L(\bx)$,
where 
\[
A(\bx)= \{\bu\in [0,\infty)^{d}: u_{j}> x_{j} \ \text{for some} \ j=1,\dots,d\}.
\]
\begin{Remark}
\textcite[Proposition 2]{fougeres2015bias} discovers the limiting behavior of $\hL_{k}$ on the Skorokhod space. Proposition \ref{prop:L_k} below extends the result in \textcite[Proposition 2]{fougeres2015bias} by considering $\hL_{k}$ as a sequence of random elements on $\ell^{\infty}([0,T]^{d})$ and deriving its limiting behavior in the Hoffman-J\o{}rgensen sense 
\textcite{hoffmann1991stochastic}, which has certain advantages (see, e.g., \textcite[p. 1598]{bucher2014uniform}).
\end{Remark}

\begin{Prop}[{cf. \textcite[Proposition 2]{fougeres2015bias}}]\label{prop:L_k}
\changeN{Suppose that Assumptions \ref{assump:firorder}, \ref{assump:secorder},
\ref{assump:thiorder}, \ref{assump:negativity}, \ref{assump:smoothness}, and \ref{assump:kspeed} hold.} Then for all $T>0$, in $\ell^{\infty}([0,T]^{d})$,
\[
\bigg\{\sqrt{k}\Big(\hL_{k}(\bx)-L(\bx)-
\alpha(n/k)M(\bx)
\Big)\bigg\}_{\bx\in [0,T]^{d}}\Rightarrow \bigg\{Z_{L}(\bx)\bigg\}_{\bx\in [0,T]^{d}},
\]
where $Z_{L}$ is defined in \eqref{eq:ZL}.
\end{Prop}
\begin{Remark}
While \textcite[Proposition 2]{fougeres2015bias} and Proposition \ref{prop:L_k} characterize the asymptotic property of $\hL_{k}(\bx)$ uniformly with respect to $\bx\in [0,T]^{d}$, Proposition \ref{prop:L_ka} below describes the limiting behavior of $\hL_{ka}(\bx)$. Compared to \textcite[Equation (12)]{fougeres2015bias}, the convergence of $\hL_{ka}(\bx)$ in Proposition \ref{prop:L_ka} is uniform in not only $\bx\in [0,T]^{d}$ but also $a\in [a_{\wedge}, a_{\vee}]$ for any $0<a_{\wedge}\leq a_{\vee}<\infty$. This uniformity in $a\in [a_{\wedge}, a_{\vee}]$ plays a key role in the proof of the consistency of the penalized estimator of $\rho$ in Theorem \ref{theo:consist} below.  
\end{Remark}

\begin{Prop}\label{prop:L_ka}
\changeN{Suppose that Assumptions \ref{assump:firorder}, \ref{assump:secorder}, \ref{assump:thiorder}, \ref{assump:negativity}, \ref{assump:smoothness}, and \ref{assump:kspeed} hold.}
Then for all $T>0$ and all $0<a_{\wedge}\leq a_{\vee}<\infty$, in $\ell^{\infty}([0,T]^{d}\times [a_{\wedge}, a_{\vee}])$,
\[
\bigg\{\sqrt{k}\Big(\hL_{ka}(\bx)-L(\bx)-
a^{-1} \alpha(n/k)M(a \bx)
\Big)\bigg\}_{\bx\in [0,T]^{d},a\in [a_{\wedge}, a_{\vee}]}\Rightarrow \bigg\{a^{-1}Z_{L}(a\bx)\bigg\}_{\bx\in [0,T]^{d}, a\in [a_{\wedge}, a_{\vee}]}.
\]
\end{Prop}
\noindent
By \eqref{eq:regv} and, e.g., \textcite[Equation (12)]{fougeres2015bias} and Proposition \ref{prop:L_ka}, intuitively one has 
\begin{equation}\label{eq:L_ka}
    \hL_{ka}(\bx)\approx L(\bx) + a^{-\rho}\alpha(n/k)M(\bx) + \text{error},
\end{equation}
Inspired by \textcite[Equation (6)]{beirlant2016bias}, after letting $i=ka$ and plugging in $a=i/k$, we could rewrite \eqref{eq:L_ka} into a regression equation, where $\hL_{i}(\bx)$ is the response, $(i/k)^{-\rho}$ is the predictor, $L(\bx)$ is the intercept, and $\alpha(n/k)M(\bx)$ is the slope:
\begin{equation}\label{eq:L_i}
\hL_{i}(\bx)\approx L(\bx) + \alpha(n/k)M(\bx)(i/k)^{-\rho} + \text{error}.
\end{equation}
\textcite[Equation (6)]{beirlant2016bias} plug in an estimated $\rho$ into \eqref{eq:L_i} and run a linear least-squares regression to estimate the intercept $L(\bx)$. Instead, in light of the Block Maxima second-order parameter estimator in \textcite[Equation (3.10)]{zou2021multiple}, we can run a non-linear least-squares regression with \eqref{eq:L_i} to estimate $\rho$: specifically, we could potentially find the $r$ that minimizes $\tRSS(b_{0},b_{1},r;\bx)$ defined in \eqref{eq:nPRSS} below. To discourage this estimate from getting too close to zero, in \eqref{eq:PRSS} we add a penalty term to $\tRSS(b_{0},b_{1},r;\bx)$. The value $r$ that minimizes $\tRSS(b_{0},b_{1},r;\bx)$ plus the penalty term becomes the penalized estimator of $\rho$, which is formalized in Section \ref{sect:pen} below.

\subsection{Penalized estimator of $\rho$}\label{sect:pen}

\begin{Def}(\changeN{Penalized estimator of $\rho$}) Suppose that  $k^{(\rho)}=k_{n}^{(\rho)}$ is a sequence of positive real numbers that can potentially differ from $k=k_{n}$. Suppose that $0<a_{\wedge}<a_{\vee}<\infty$,  $A=[a_{\wedge},a_{\vee}]$, {$M_{n}=\{\floor{k^{(\rho)}a}: a\in A\}$}, $K'<K''<0$, $\eta\geq 0$,  $\cX_{\rho}\subset [0,\infty)^{d}$, and $\{w_{n,i}, i\in M_{n}\}$ are weights satisfying Assumption \ref{assump:weight} below. Let the non-penalized form of residual sum of squares be defined as 
\begin{equation}\label{eq:nPRSS}
    \tRSS(b_{0},b_{1},r;\bx)=\sum_{i\in M_{n}}w_{n,i}\Big[\hL_{i}(\bx)-b_{0}-b_{1}(i/{k^{(\rho)}})^{-r}\Big]^{2}.
\end{equation}
Further, let $\hRSS_\eta(b_{0},b_{1},r;\bx)$, the penalized form of residual sum of squares, be defined as 
\begin{equation}\label{eq:PRSS}
\hRSS_\eta(b_{0},b_{1},r;\bx)
=\tRSS(b_{0},b_{1},r;\bx) + \frac{\eta}{|r|} {\min_{a_{0}, a_{1} \in \bbR, K'\leq \kappa\leq K''}\tRSS(a_{0},a_{1},\kappa;\bx)}.
\end{equation}
Finally, let $\hrho_{\cX_{\rho}}^{\mathrm{pen,agg}}$, the penalized estimator of $\rho$, be defined as 
\begin{equation}\label{eq:reg_rho}
\begin{aligned}
(\hb_0(\bx), \hb_1(\bx),  \hrho^{\mathrm{pen}}(\bx)) &= \argmin_{b_0,b_1 \in \bbR, K'\leq r\leq K''} \hRSS_\eta(b_{0},b_{1},r;\bx),\\
\hrho_{\cX_{\rho}}^{\mathrm{pen,agg}} &= \frac{1}{|\cX_{\rho}|} \sum_{\bx \in \cX_{\rho}} \hrho^{\mathrm{pen}}(\bx),
\end{aligned} 
\end{equation}
where a choice of tuning parameters $w_{n,i}$, $M_{n}$, $K'$, $K''$, $\eta$, and $\cX_{\rho}$ is specified in Section \ref{sect:tune}. 
\end{Def}


\begin{Assump}[Properties of weight $w_{n,i}$]\label{assump:weight}
Recall that in the definition above of the \changeN{penalized} estimator, $0<a_{\wedge}<a_{\vee}<\infty$ and $A=[a_{\wedge},a_{\vee}]$. Now assume that uniformly over $A$,  \changeN{${k^{(\rho)}}w_{n,\floor{k^{(\rho)}a}}\to f(a)$} for some positive, continuous $f$ on $A$; {in addition, assume that $\sum_{i\in M_{n}}w_{n,i}=1$. }
\end{Assump}

\begin{Theo}[Uniform consistency of $\hrho^{\mathrm{pen}}$]\label{theo:consist}
Suppose that Assumptions \ref{assump:firorder}, \ref{assump:secorder}, \ref{assump:thiorder}, \ref{assump:smoothness}, and \ref{assump:weight} hold. Suppose that Assumption \ref{assump:negativity} holds, and $\rho\in [K', K'']$. Suppose that Assumption \ref{assump:kspeed} is met by $k^{(\rho)}=k_{n}^{(\rho)}$ instead of $k=k_{n}$. Then, for any fixed $\eta\geq 0$, any $T>0$, and any compact set $\cX\subset \{\bx\in [0,T]^{d}: M(\bx)\neq 0\}$,
\[
\sup_{\bx\in \cX}|\hrho^{\mathrm{pen}}(\bx)-\rho|=\op{1}.
\]
\end{Theo}
\begin{Coro}[Consistency of $\hrho_{\cX_{\rho}}^{\mathrm{pen,agg}}$]\label{coro:consist}
Suppose that the assumptions in Theorem \ref{theo:consist} hold. Suppose that there exist some $T>0$ and some compact set $\cX$ such that $\cX_{\rho}\subset\cX \subset \{\bx\in [0,T]^{d}: M(\bx)\neq 0\}$. Then for all fixed $\eta\geq 0$,
\[
\hrho_{\cX_{\rho}}^{\mathrm{pen,agg}}-\rho = \op{1}.
\]
\end{Coro}
\subsection{Estimation of $L$}\label{sect:estL}
With some misuse of notations, $\rho$ below indicates both the true second-order parameter and the argument in the functions. Now we plug in the penalized estimator $\hrho_{\cX_{\rho}}^{\mathrm{pen,agg}}$ into the existing bias-corrected estimators below. 
By \textcite[Theorem 3]{fougeres2015bias} and \textcite[Theorem 2]{beirlant2016bias}, these bias-corrected estimators can successfully reduce the bias to a smaller order when equipped with the true second-order parameter $\rho$. Hence, to illustrate the theoretical performance of the penalized estimator $\hrho_{\cX_{\rho}}^{\mathrm{pen,agg}}$, it may suffice to bound the difference between the biased-corrected estimator equipped with the true second-order parameter $\rho$ and those equipped with the penalized estimator $\hrho_{\cX_{\rho}}^{\mathrm{pen,agg}}$. For simplicity, we provide this upper bound only for the \textcite{fougeres2015bias} dot estimator $\mL_{k,a}(\bx,\rho)$; see Proposition \ref{prop:PlugInErrorDot}. We conjecture that similar upper bounds apply to other bias-corrected estimators below as well. 

\subsubsection{Non-bias-corrected estimator of $L$}
The non-bias-corrected estimator, $\hL_{k}(\bx)$, by \textcite{huang1992statistics}, is defined in \eqref{eq:hatL}.

\subsubsection{\textcite{beirlant2016bias} estimator of $L$}
The \textcite{beirlant2016bias} estimator, $\bar{L}_{k,\bar{k}}(\bx,\rho)$, as a bias-corrected estimator of $L$, is defined by $a_{j,k} = j/(k+1)$, $K(t)=(\tau+1)t^{\tau}\ind_{\{t\in (0,1)\}}$, $K_{B}(t)=(\tau_{B}+1)t^{\tau_{B}}\ind_{\{t\in (0,1)\}}$, and
\begin{equation}\label{eq:Bei_L}
\begin{aligned}
    \tL_{k}(\bx)&= \frac{1}{k}\sum_{j=1}^{k}K(a_{j,k})\hL_{k,a_{j,k}}(\bx),\\
    \talpha_{k}(\bx,\rho)&=\frac{\sum_{j=1}^{k}\sum_{l=1}^{k}K_{B}(a_{j,k})K_{B}(a_{l,k})\bigg(a_{j,k}^{-\rho}-a_{l,k}^{-\rho}\bigg)\hL_{k,a_{j,k}}(\bx)}{\sum_{j=1}^{k}\sum_{l=1}^{k}K_{B}(a_{j,k})K_{B}(a_{l,k})\bigg(a_{j,k}^{-\rho}-a_{l,k}^{-\rho}\bigg)a_{j,k}^{-\rho}},\\
    \bar{L}_{k,\bar{k}}(\bx,\rho)&=\frac{\tL_{k}(\bx)-(\bar{k}/k)^{\rho}\talpha_{\bar{k}}(\bx,\rho)\frac{1}{k}\sum_{j=1}^{k}K(a_{j,k})a_{j,k}^{-\rho}}{\frac{1}{k}\sum_{j=1}^{k}K(a_{j,k})},
\end{aligned}
\end{equation}
where a choice of tuning parameters $\bar{k}$, $\tau$, and $\tau_{B}$ is specified in Section \ref{sect:tune}, and $\hL_{ka}(\bx)$ is defined in \eqref{eq:hatL}.

\subsubsection{\textcite{fougeres2015bias} dot estimator of $L$}
The \textcite{fougeres2015bias} dot estimator $\mL_{k,a}(\bx,\rho)$, as a bias-corrected estimator of $L$, is defined in \eqref{eq:Fou_dot_L}, \changeN{where a choice of tuning parameter $a$ is specified in Section \ref{sect:tune},} and $\hL_{ka}(\bx)$ is defined in \eqref{eq:hatL}.

\begin{Prop}\label{prop:PlugInErrorDot}
Recall that $\rho$ indicates the true second-order parameter. Suppose the assumptions in Corollary \ref{coro:consist} hold. Suppose that Assumption \ref{assump:kspeed} also holds for $k=k_{n}$ in addition to  $k^{(\rho)}=k_{n}^{(\rho)}$. Suppose $a>0$. Then
\[
\sup_{\bx\in [0,T]^{d}}\sqrt{k}\left|\mL_{k,a}(\bx,\hrho_{\cX_{\rho}}^{\mathrm{pen,agg}})-\mL_{k,a}(\bx,\rho)\right| =  \Op{\sqrt{k}\alpha(n/k)(\hrho_{\cX_{\rho}}^{\mathrm{pen,agg}}-\rho)}+\op{1}.
\]

\end{Prop}


\subsubsection{\textcite{fougeres2015bias} dot aggregated estimator of $L$}
The \textcite{fougeres2015bias} dot aggregated estimator $\mL_{\cK,a}^{\text{agg}}(\bx,\rho)$, as a bias-corrected estimator of $L$, is defined by
\begin{equation}\label{eq:Fou_dot_agg_L}
\begin{aligned}
    \mL_{\cK,a}^{\text{agg}}(\bx,\rho)=\text{Median}(\{ \mL_{k,a}(\bx,\rho), \ k\in \cK\}).
\end{aligned}
\end{equation}
\changeN{where a choice of tuning parameters $\cK$ and $a$ is specified in Section \ref{sect:tune},} and  $\mL_{k,a}(\bx,\rho)$ is defined in \eqref{eq:Fou_dot_L}.






\newpage 

\section{Simulation}\label{sect:simul}

\subsection{Data Generating Processes}

\changeN{The collection of data generating processes in our simulation (Table \ref{table:dgp}) contains those used in \textcite{beirlant2016bias} but also includes some extra potentially interesting cases.

Specifically, \textcite{beirlant2016bias} considers in its simulation data generating processes ``Cauchy", ``$t_2$",  ``BPII(3)", ``Symmetric logistic", ``Archimax logistic", and ``Archimax mixed", where the choices of parameters for these data generating processes are the same as those in Table \ref{table:dgp}. In these data generating processes, the true second-order parameters indeed all have their absolute values $|\rho|\geq 1$. On the other hand, $\rho$ with a smaller absolute value poses more serious challenges to the bias-corrected estimators; intuitively, when $\rho$ has a smaller absolute value, the bias is small, so the bias-corrected estimator does have too much to correct; technically, the error in the estimation of $\rho$ usually has an order of $1/(\sqrt{k}\alpha(n/k))$ (see \textcite[Proposition 6]{fougeres2015bias} and \textcite[Proposition 1]{beirlant2016bias}), which, since $\alpha(n/k)$ has an approximate order of $(n/k)^{-|\rho|}$, tends to increase when $\rho$ has a smaller absolute value. By \textcite[pp. 911-912]{fougeres2015bias}, $t$-copula with degrees of freedom $4$ and $6$ has $|\rho|=1/2$ and $|\rho|=1/3$, respectively. Hence, to analyze the bias-corrected estimator when $\rho$ has a smaller absolute value, we include $t_{4}$ and $t_6$ in our data generating processes. }

\begin{table}[H]
\begin{tabular}{|c|c|}
\hline
Data Generating Process & Description                                                                       \\ \hline
Cauchy                  & $t$-copula with degrees of freedom 1 and correlation coefficient $\theta=0$   \\
$t_{2}$                      & $t$-copula with degrees of freedom 2 and correlation coefficient $\theta=0.5$ \\
$t_{4}$                      & $t$-copula with degrees of freedom 4 and correlation coefficient $\theta=0.5$ \\
$t_{6}$                      & $t$-copula with degrees of freedom 6 and correlation coefficient $\theta=0.5$ \\
BPII(3)                 & bivariate Pareto of type II distribution with $\beta=3$                            \\
Symmetric logistic                & bivariate symmetric logistic distribution with $s=1/3$                           \\
Archimax logistic                & Archimax model with logistic generator $L(x,y)=(x^{2}+y^{2})^{1/2}$                          \\
Archimax mixed                & Archimax model with mixed generator $L(x,y)=(x^{2}+y^{2}+xy)/(x+y)$                          \\
\hline
\end{tabular}
\caption{List of Data Generating Processes}
\label{table:dgp}
\end{table}

\subsection{Methodologies}
Table \ref{table:est} includes the names of the estimators, the notations of the estimators, and the colors corresponding to the estimators in Figure \ref{fig:multipoint_1} and Figure \ref{fig:multipoint_2}. Notice that we only plugged the penalized estimator of $\rho$ to the \textcite{fougeres2015bias} dot aggregated estimator of $L$ and the \textcite{beirlant2016bias} estimator of $L$, because the \textcite{fougeres2015bias} dot aggregated estimator appears to be the most competitive estimator in \textcite{fougeres2015bias} (see \textcite[pp. 917-919]{fougeres2015bias}), and the \textcite{beirlant2016bias} estimator of $L$ is the only estimator used in \textcite{beirlant2016bias} and \textcite{goegebeur2017kernel}. 

Under the \textcite{fougeres2015bias} dot aggregated estimator of $L$, the penalized estimator of $\rho$ is compared with the \textcite{fougeres2015bias} aggregated estimator of $\rho$; under the \textcite{beirlant2016bias} estimator of $L$, the penalized estimator of $\rho$ is compared with the \textcite{beirlant2016bias} estimator of $\rho$ and the \textcite{goegebeur2017kernel} estimator of $\rho$. We also include the result of the \textcite{fougeres2015bias} dot (non-aggregated) estimator of $L$ and the non-bias-corrected estimator of $L$ for comparison. Notice that we have also simulated the \textcite[Equation (17)]{fougeres2015bias} tilde estimator of $L$ but choose not to present the result since this tilde estimator performs significantly worse than the \textcite{fougeres2015bias} dot estimator; this under-performance of the tilde estimator has also been pointed out on \textcite[p. 917]{fougeres2015bias}.

\begin{table}[H]
\begin{tabular}{|llll|}
\hline
Estimator of $L$   & Estimator of $\rho$   & Notation                                                  & Color in graphs \\ \hline
Non-bias-corrected & N/A                   & $\hL_{k}(\bx)$                                            & Black              \\
\textcite{fougeres2015bias} dot        & \textcite{fougeres2015bias}            & $\mL_{k,a}(\bx,\mrho_{\bar{k}}(\bx))$                 & Purple             \\
\textcite{fougeres2015bias} dot        & \textcite{fougeres2015bias} aggregated & $\mL_{k,a}(\bx,\mrho_{\bar{k}}^{\text{agg}})$                    & Red                \\
\textcite{fougeres2015bias} dot aggregated        & \textcite{fougeres2015bias} aggregated & 
$\mL_{\cK,a}^{\text{agg}}(\bx,\mrho_{\bar{k}}^{\text{agg}})$                  & Orange                \\
\textcite{fougeres2015bias} dot aggregated        & \changeN{Penalized}  & 
$\mL_{\cK,a}^{\text{agg}}(\bx,\hrho_{\cX_{\rho}}^{\mathrm{pen,agg}})$                  & Dashed-Orange                \\
\textcite{beirlant2016bias}         & \textcite{beirlant2016bias}            & $\bar{L}_{k,\bar{k}}(\bx,\trho_{\bar{k}}(\bx))$       & Blue               \\
\textcite{beirlant2016bias}         & \textcite{goegebeur2017kernel}      & $\bar{L}_{k,\bar{k}}(\bx,\trho_{\bar{k},\xi_{1},\xi_{2}}(\bx))$ & Dotted-Blue           \\ 
\textcite{beirlant2016bias}         & \changeN{Penalized}      & $\bar{L}_{k,\bar{k}}(\bx,\hrho_{\cX_{\rho}}^{\mathrm{pen,agg}})$ & Dashed-Blue               \\ 
\hline

\end{tabular}
\caption{List of Estimators \changeN{of $L$}}
\label{table:est}
\end{table}

\subsubsection{\textcite{fougeres2015bias} estimator of $\rho$}
The \textcite{fougeres2015bias} estimator $\mrho_{\bar{k}}(\bx)$ is defined by 
\begin{equation}\label{eq:Fou_rho}
\begin{aligned}
    \hDelta_{\bar{k},a}(\bx)&=a^{-1}\hL_{\bar{k}}(a\bx)-\hL_{\bar{k}}(\bx),\\
    \mrho_{\bar{k}}(\bx)&=\bigg(1-\frac{1}{\log r}\log \Big|\frac{\hDelta_{\bar{k},a}(r\bx)}{\hDelta_{\bar{k},a}(\bx)}\Big|\bigg)\wedge 0.
\end{aligned}
\end{equation}
\changeN{where a choice of tuning parameters $\bar{k}$, $a$, $r$, and $\bx$ is specified in Section \ref{sect:tune}}.

\subsubsection{\textcite{beirlant2016bias} estimator of $\rho$}
The \textcite{beirlant2016bias} estimator $\trho_{\bar{k}}(\bx)$ is defined by
\begin{equation}\label{eq:Bei_rho}
\begin{aligned}
    \tDelta_{\bar{k},a}(\bx)&=a^{-1}\tL_{\bar{k}}(a\bx)-\tL_{\bar{k}}(\bx),\\
    \trho_{\bar{k}}(\bx)&=\bigg(1-\frac{1}{\log r}\log \Big|\frac{\tDelta_{\bar{k},a}(r\bx)}{\tDelta_{\bar{k},a}(\bx)}\Big|\bigg)\wedge 0,
\end{aligned}
\end{equation}
\changeN{where a choice of tuning parameters $\bar{k}$, $a$, $r$, and $\bx$ is specified in Section \ref{sect:tune},} and $\tL_{\bar{k}}(\bx)$ is defined in \eqref{eq:Bei_L}.

\subsubsection{\textcite{goegebeur2017kernel} estimator of $\rho$}
The \textcite{goegebeur2017kernel} estimator $\trho_{\bar{k},\xi_{1},\xi_{2}}(\bx)$ is defined by $K(t) =(\tau+1)t^{\tau}\ind_{\{t\in [0,1]\}}$, and
\begin{equation}\label{eq:Qin_rho}
\begin{aligned}
    \tL_{k,\xi}(\bx)&= \frac{1}{k}\sum_{j=1}^{k}K(a_{j,k})\hL_{k,a_{j,k}}^{\xi}(\bx),
    \\
    \tDelta_{\bar{k},a,\xi_{1},\xi_{2}}(\bx)&=\big[a^{-\xi_{1}}\tL_{\bar{k},\xi_{1}}(a\bx)\big]^{1/\xi_{1}}-\big[\tL_{\bar{k},\xi_{2}}(\bx)\big]^{1/\xi_{2}},\\
    \trho_{\bar{k},\xi_{1},\xi_{2}}(\bx)&=\bigg(1-\frac{1}{\log r}\log \Big|\frac{\tDelta_{\bar{k},a,\xi_{1},\xi_{2}}(r\bx)}{\tDelta_{\bar{k},a,\xi_{1},\xi_{2}}(\bx)}\Big|\bigg)\wedge 0,
\end{aligned}
\end{equation}
\changeN{where a choice of tuning parameters $\bar{k}$, $\xi_{1}$, $\xi_{2}$, $\tau$, $a$, $r$, and $\bx$ is specified in Section \ref{sect:tune},} and $\hL_{ka}(\bx)$ is defined in $\eqref{eq:hatL}$.

\subsubsection{Aggregated \textcite{fougeres2015bias} estimator of $\rho$}
The Aggregated \textcite{fougeres2015bias} estimator $\mrho_{\bar{k}}^{\text{agg}}$ is defined by
\begin{equation}\label{eq:agg_Fou_rho}
\mrho_{\bar{k}}^{\text{agg}} = \frac{1}{\cX_{\rho}}\sum_{\bx\in \cX_{\rho}}\mrho_{\bar{k}}(\bx),
\end{equation}
where $\mrho_{\bar{k}}(\bx)$ is defined in \eqref{eq:Fou_rho}, and 
\changeN{a choice of tuning parameters $\bar{k}$, $\cX_{\rho}$, $a$, and $r$ is specified in Section \ref{sect:tune}}.

\subsection{Tuning Parameters}\label{sect:tune}


\changeN{We set the tuning parameters according to the suggestions by \textcite{fougeres2015bias}, \textcite{beirlant2016bias}, and \textcite{goegebeur2017kernel}. The specific settings are detailed below.}

\subsubsection{\changeN{Penalized} estimator of $\rho$}
When generating $\hrho_{\cX_{\rho}}^{\mathrm{pen,agg}}$, \changeN{in \eqref{eq:reg_rho}} we let $w_{n,i} \ \propto \ i$ and $M_{n}=\{50,100,\dots,1000$\} for all $k$, and let $K'= -4$, $K''=-0.1$, $\eta = 0.5$, and $\cX_{\rho}=\{(0.3,0.3), (0.35,0.35),\dots, (0.7,0.7)\}$. In the minimization procedure, we search for the minimizer $\rho$ on the grid of $\{-4, -3.9, \dots, -0.1\}$.

\subsubsection{\textcite{fougeres2015bias} estimator of $\rho$}
When generating $\mrho_{\bar{k}}(\bx)$, \changeN{in \eqref{eq:Fou_rho}} we let $\bar{k}=990$, $a=r=0.4$, \changeN{and $\bx$ be the same $\bx$ at which we estimate $L(\bx)$.} If $\mrho_{\bar{k}}(\bx)>-0.1$, we set $\mrho_{\bar{k}}(\bx)=-1$.

\subsubsection{\textcite{beirlant2016bias} estimator of $\rho$}
When generating $\trho_{\bar{k}}(\bx)$, \changeN{in \eqref{eq:Bei_rho}} we let $\bar{k}=990$, $a=r=0.4$, \changeN{and $\bx$ be the same $\bx$ at which we estimate $L(\bx)$}. If $\trho_{\bar{k}}(\bx)>-0.1$,  we set $\trho_{\bar{k}}(\bx)=-1$. 

\subsubsection{\textcite{goegebeur2017kernel} estimator of $\rho$}
When generating $\trho_{\bar{k},\xi_{1},\xi_{2}}(\bx)$, \changeN{in \eqref{eq:Qin_rho}} we let $\bar{k}=990$, $\xi_{1}=\xi_{2}=4$, $\tau=10$, $a=r=0.4$, \changeN{and $\bx$ be the same $\bx$ at which we estimate $L(\bx)$}. If $\trho_{\bar{k},\xi_{1},\xi_{2}}(\bx)>-0.1$,  we set $\trho_{\bar{k},\xi_{1},\xi_{2}}(\bx)=-1$. 

\subsubsection{Aggregated \textcite{fougeres2015bias} estimator of $\rho$}
When generating $\mrho_{\bar{k}}^{\text{agg}}$, \changeN{in \eqref{eq:agg_Fou_rho}} we let $\bar{k}=990$, $\cX_{\rho}=\{(0.3,0.3), (0.35,0.35),\dots, (0.7,0.7)\}$, and $a=r=0.4$. In the aggregation, whenever $\mrho_{\bar{k}}(\bx)>-0.1$, we set $\mrho_{\bar{k}}(\bx)=-1$.

\subsubsection{\textcite{beirlant2016bias} estimator of $L$}
When generating $\bar{L}_{k,\bar{k}}(\bx,\rho)$, \changeN{in \eqref{eq:Bei_L}} we let $\bar{k}=990$, $\tau=5$, $\tau_{B}=0.5$, and truncate $\bar{L}_{k,\bar{k}}(\bx,\rho)$ so that $\max (x_{1},x_{2})\leq \bar{L}_{k,\bar{k}}((x_{1},x_{2}),\rho)\leq x_{1}+x_{2}$.

\subsubsection{\textcite{fougeres2015bias} dot estimator of $L$}
When generating $\mL_{k,a}(\bx,\rho)$, \changeN{in \eqref{eq:Fou_dot_L}} we let $a=0.4$ and truncate $\mL_{k,a}(\bx,\rho)$ so that $\max (x_{1},x_{2})\leq \mL_{k,a}((x_{1},x_{2}),\rho)\leq x_{1}+x_{2}$.

\subsubsection{\textcite{fougeres2015bias} dot aggregated estimator of $L$}
When generating $\mL_{\cK,a}^{\text{agg}}(\bx,\rho)$, \changeN{in \eqref{eq:Fou_dot_agg_L}} we let $\cK=\{1,51,101,\dots, 951\}$, and $a=0.4$. Before the aggregation, we truncate $\mL_{k,a}(\bx,\rho)$ so that $\max (x_{1},x_{2})\leq \mL_{k,a}((x_{1},x_{2}),\rho)\leq x_{1}+x_{2}$.

\subsection{Result}
\begin{figure}[H]   
\centering
\subfigure{
\includegraphics[width=.280\textwidth]{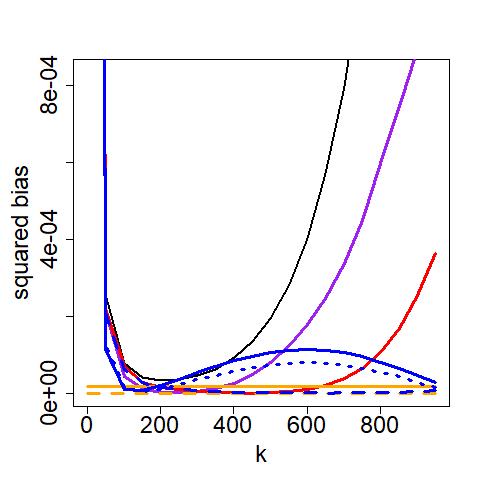}
}
\subfigure{
\includegraphics[width=.280\textwidth]{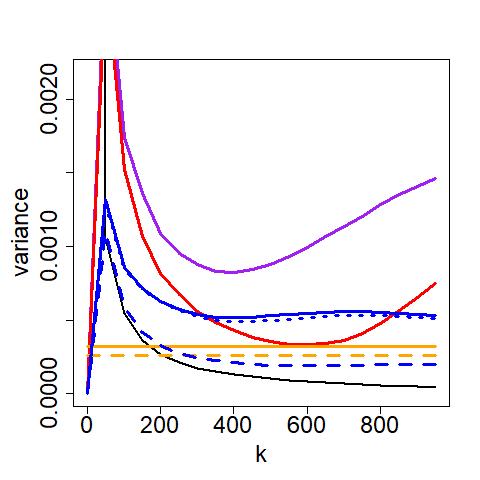}
}
\subfigure{
\includegraphics[width=.280\textwidth]{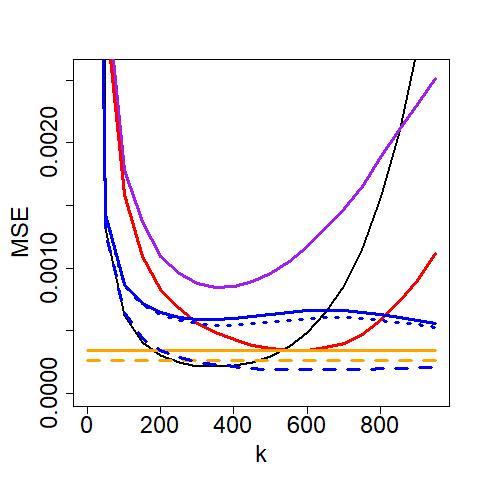}
}
\\
\subfigure{
\includegraphics[width=.280\textwidth]{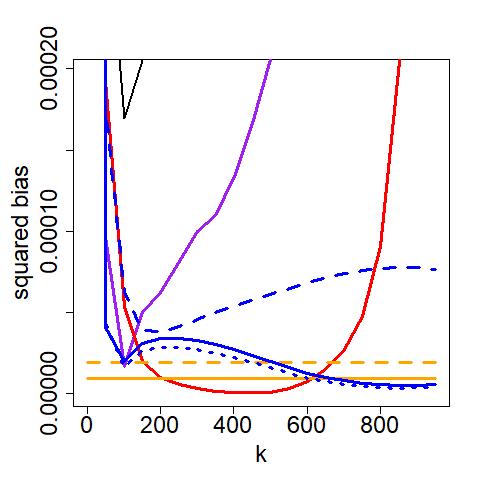}
}
\subfigure{
\includegraphics[width=.280\textwidth]{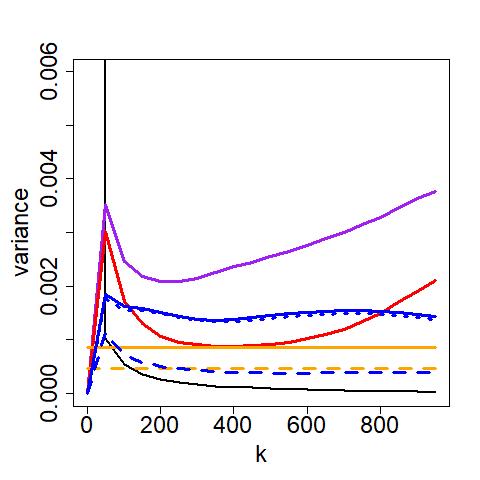}
}
\subfigure{
\includegraphics[width=.280\textwidth]{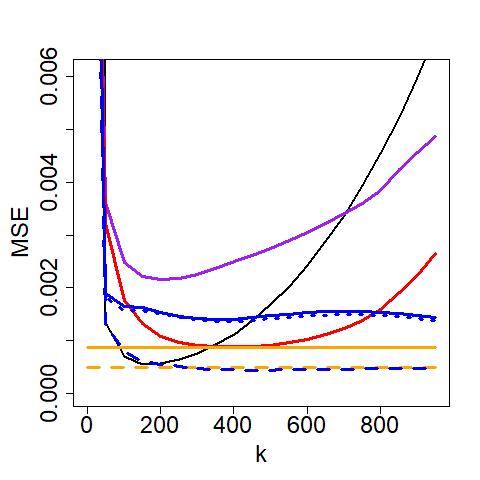}
}
\\
\subfigure{
\includegraphics[width=.280\textwidth]{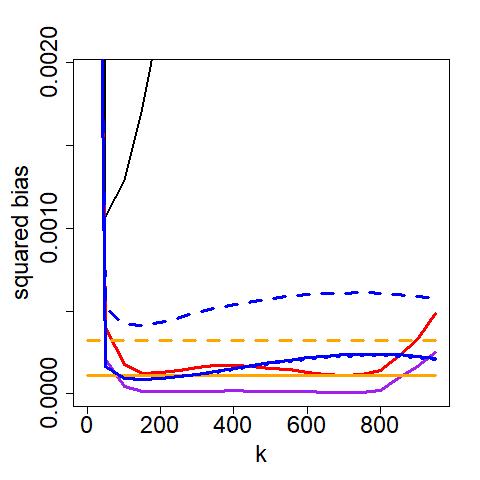}
}
\subfigure{
\includegraphics[width=.280\textwidth]{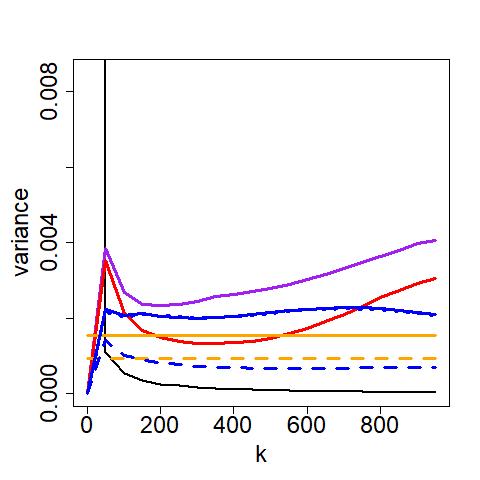}
}
\subfigure{
\includegraphics[width=.280\textwidth]{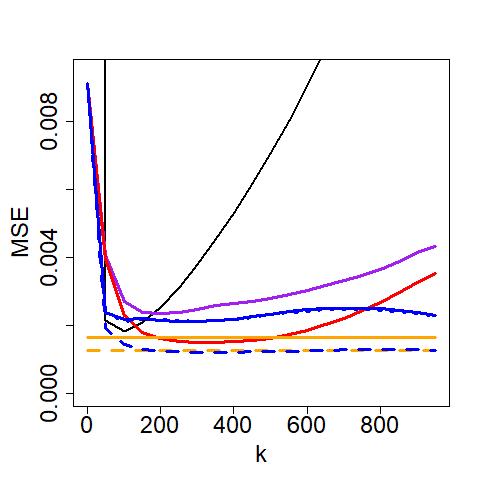}
}
\\
\subfigure{
\includegraphics[width=.280\textwidth]{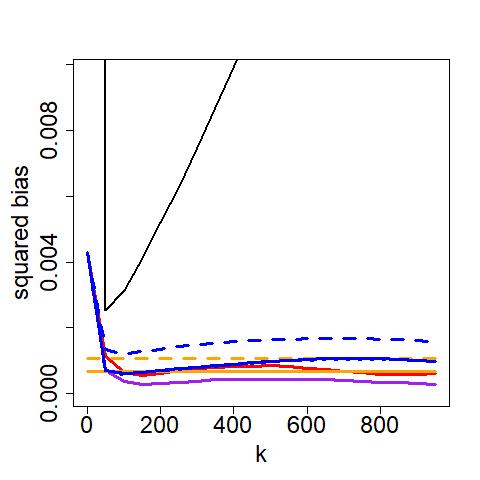}
}
\subfigure{
\includegraphics[width=.280\textwidth]{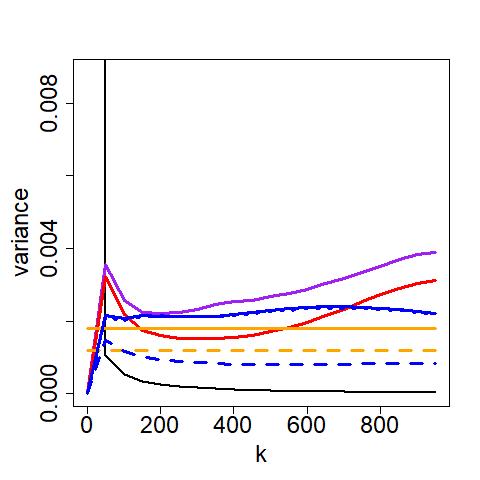}
}
\subfigure{
\includegraphics[width=.280\textwidth]{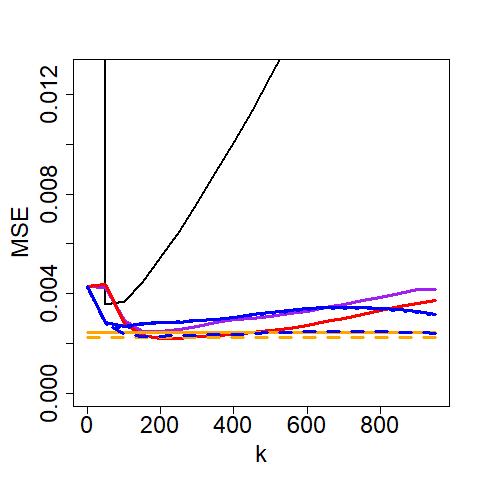}
}
\caption{Squared Bias (left), Variance (Middle), and MSE (Right) of estimators of $L$ in Table \ref{table:est} on data generating processes Cauchy (1st row), $t_{2}$ (2nd row), $t_{4}$ (3rd row), and $t_{6}$ (4th row) in Table \ref{table:dgp}}
\label{fig:multipoint_1}
\end{figure}

\begin{figure}[H]   
\centering
\subfigure{
\includegraphics[width=.280\textwidth]{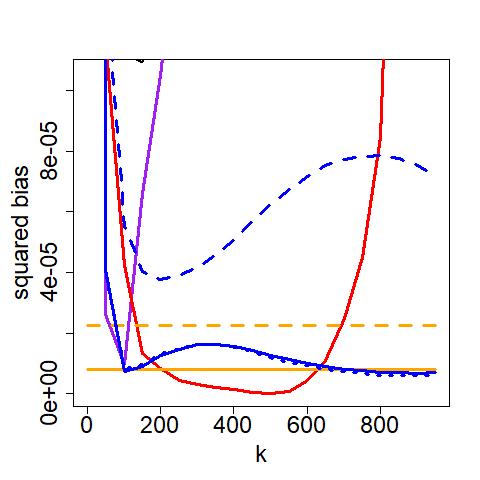}
}
\subfigure{
\includegraphics[width=.280\textwidth]{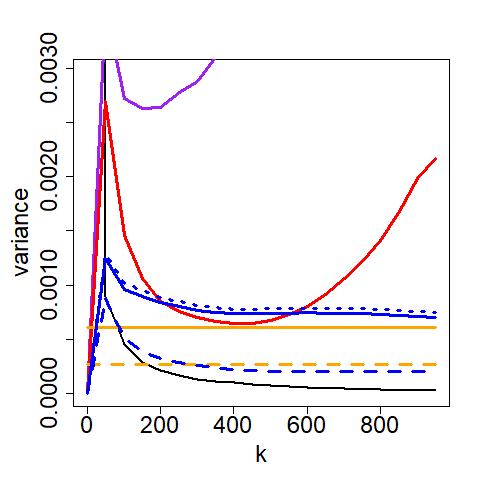}
}
\subfigure{
\includegraphics[width=.280\textwidth]{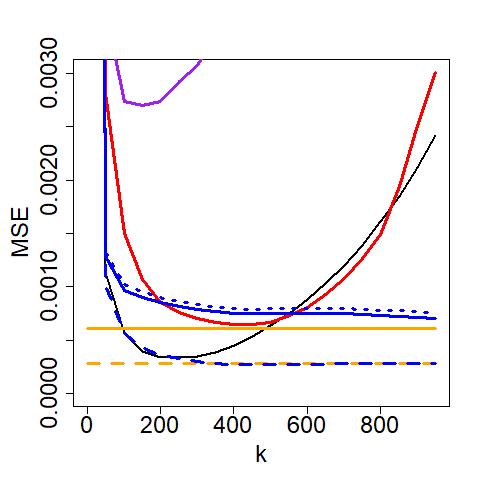}
}
\\
\subfigure{
\includegraphics[width=.280\textwidth]{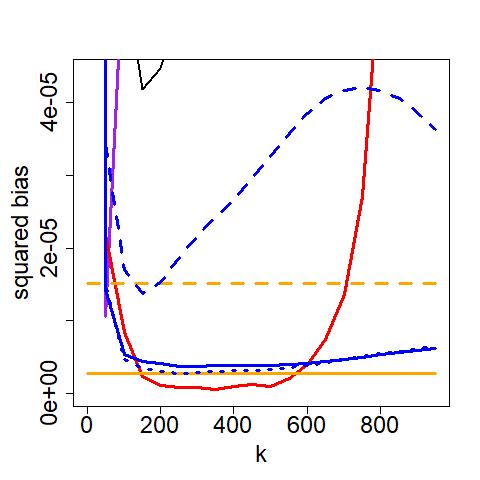}
}
\subfigure{
\includegraphics[width=.280\textwidth]{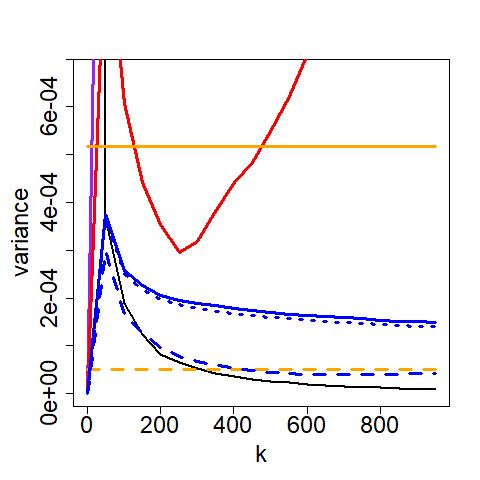}
}
\subfigure{
\includegraphics[width=.280\textwidth]{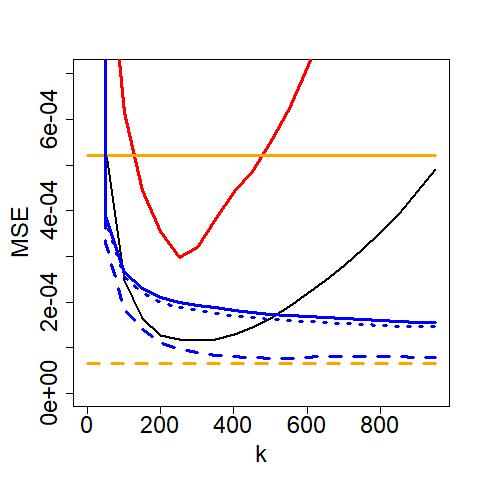}
}
\\
\subfigure{
\includegraphics[width=.280\textwidth]{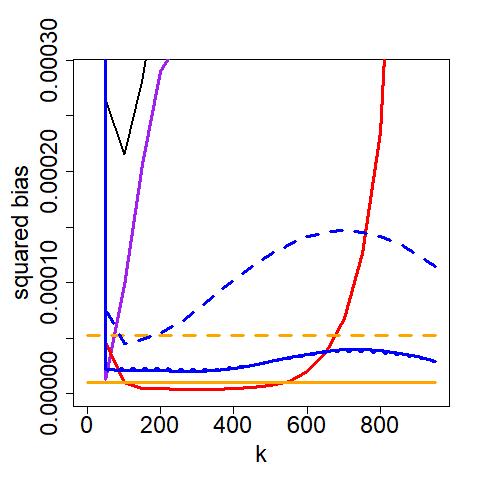}
}
\subfigure{
\includegraphics[width=.280\textwidth]{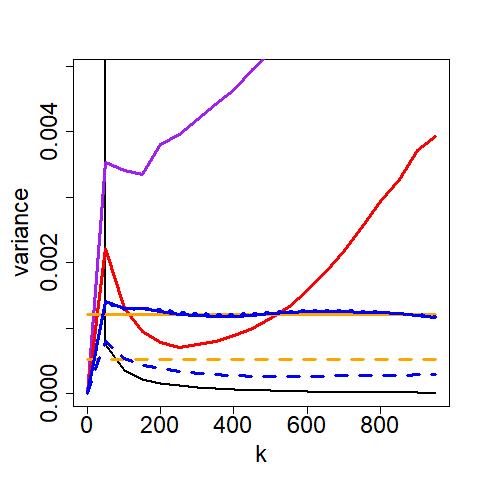}
}
\subfigure{
\includegraphics[width=.280\textwidth]{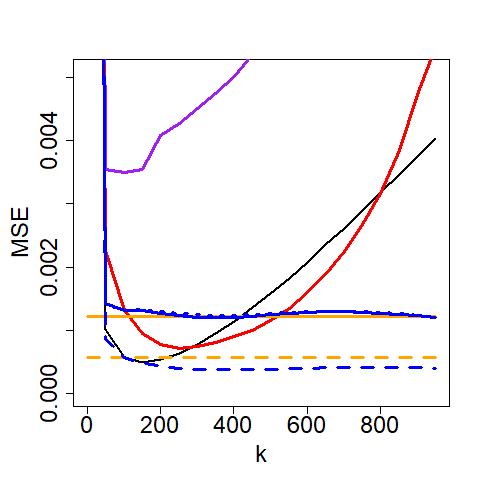}
}
\\
\subfigure{
\includegraphics[width=.280\textwidth]{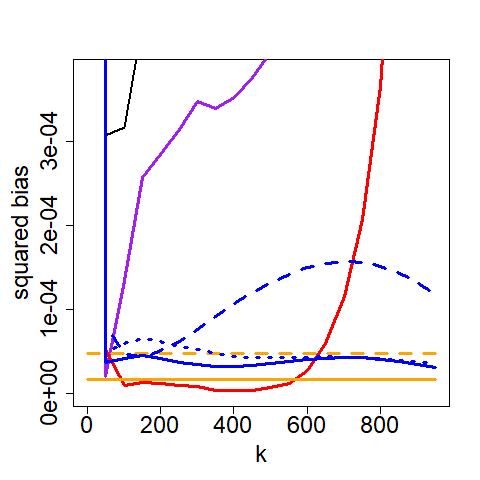}
}
\subfigure{
\includegraphics[width=.280\textwidth]{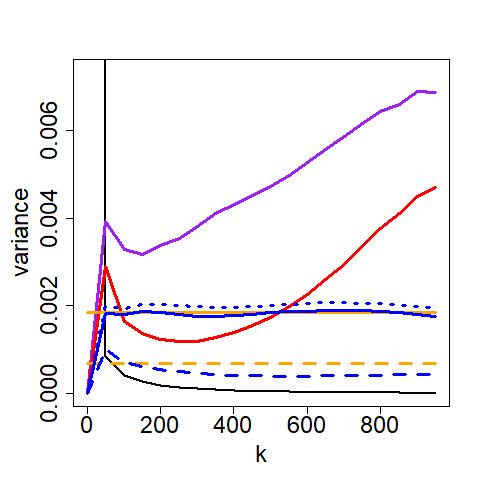}
}
\subfigure{
\includegraphics[width=.280\textwidth]{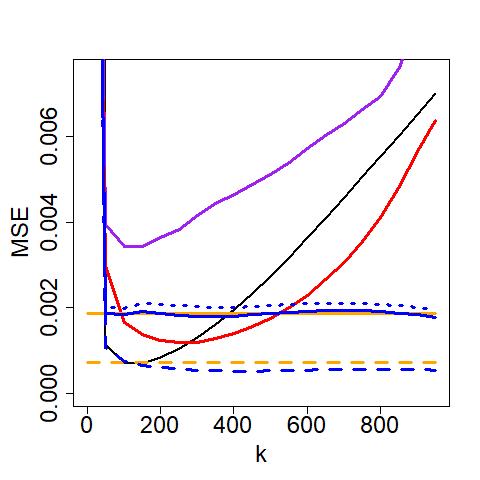}
}
\caption{Squared Bias (left), Variance (Middle), and MSE (Right) of estimators of $L$ \changeN{in Table \ref{table:est}} on data generating processes \changeN{BPII(3) (1st row), Symmetric logistic (2nd row), Archimax logistic (3rd row), and Archimax mixed (4th row)} in Table \ref{table:dgp}}
\label{fig:multipoint_2}
\end{figure}

In our simulation, we estimate
\begin{align*}
    \changeN{\text{Squared Bias}} &= \frac{1}{|\cX|}\sum_{x\in \cX}\Big[\Exp\big(\check {L}_{k}(\bx)-L(\bx)\big)\Big]^{2}\\
    \changeN{\text{Variance}} &= \frac{1}{|\cX|}\sum_{x\in \cX}\Var\big(\check {L}_{k}(\bx)\big)\\
    \text{MSE} &=\frac{1}{|\cX|}\sum_{x\in \cX} \Exp\Big[\big(\check {L}_{k}(\bx)-L(\bx)\big)^{2}\Big]
\end{align*}
with $\check {L}_{k}$ being an estimator of $L$ listed in Table \ref{table:est}, $\cX=
\{(t, 1-t), t = 0.1, 0.2, \dots, 1\}$ \changeN{identical to the points considered in \textcite{beirlant2016bias},} $k= 1, 51, 101, \dots, 951$, and $N=1000$ iterations. In each iteration, we work on a sample of size 1000 from data generating processes detailed in Table \ref{table:dgp}. For each curve in Figure \ref{fig:multipoint_1} and Figure \ref{fig:multipoint_2}, the corresponding estimator is specified in Table \ref{table:est}, and the corresponding tuning parameters are specified in Section \ref{sect:tune}.

\changeN{
Notice that due to the substantial difference among the behaviors of the estimators, in some sub-figures in Figure \ref{fig:multipoint_1} and Figure \ref{fig:multipoint_2}, the curves of some estimators, especially the non-bias-corrected estimator of $L$ (black) and the \textcite{fougeres2015bias} dot estimator of $L$ with the \textcite{fougeres2015bias} estimator of $\rho$ (purple), are out of the scales of the sub-figures; additionally, in some sub-figures, the curves of some estimators, especially the \textcite{beirlant2016bias} estimator of $L$ 
with the \textcite{beirlant2016bias} estimator of $\rho$ (blue) and the \textcite{beirlant2016bias} estimator of $L$ with the \textcite{goegebeur2017kernel} estimator of $\rho$ (dotted-blue), overlap with each other.  

On the dependence of the bias and  variance on the tuning parameter $k$, since larger $k$ indicates a smaller threshold, when $k$ increases, the bias of the non-bias-corrected estimator of $L$ (black) increases. On the other hand, since $\hL_{k}$ averages $k$ data, when $k$ increases, the variance of the non-bias-corrected estimator of $L$ (black) decreases. Notice this increase/decrease pattern of bias and variance is not shared by the \textcite{fougeres2015bias} dot aggregated estimator of $L$ with the \textcite{fougeres2015bias} aggregated estimator of $\rho$ (orange) and the \textcite{fougeres2015bias} dot aggregated estimator of $L$ with the penalized estimator of $\rho$ (dashed-orange) because the \textcite{fougeres2015bias} dot aggregated estimator of $L$ does not really depend on $k$. This increase/decrease pattern of bias and variance is not necessarily shared by other bias-corrected estimators, either, because the biases and variances of the bias-corrected estimators also involve the biases and variances from the estimation of $\rho$. 

Now we analyze the bias, variance, and MSE of estimators of $L$ equipped with the penalized estimator of $\rho$. Recall that, under the \textcite{fougeres2015bias} dot aggregated estimator of $L$, the penalized estimator of $\rho$ (dashed-orange) is compared with the \textcite{fougeres2015bias} aggregated estimator of $\rho$ (orange). Meanwhile, under the \textcite{beirlant2016bias} estimator of $L$, the penalized estimator of $\rho$ (dashed-blue) is compared with the \textcite{beirlant2016bias} estimator of $\rho$ (blue) and the \textcite{goegebeur2017kernel} estimator of $\rho$ (dotted-blue). In these comparisons, the penalized estimators of $\rho$ sacrifice slightly in the empirical bias yet significantly reduce the empirical variance. As a consequence, the penalized estimators of $\rho$ uniformly improve the empirical MSE, for almost all choices of the tuning parameter $k$ and under all data generating processes.

Finally, the two penalized estimators, namely the \textcite{fougeres2015bias} dot aggregated estimator of $L$ with the penalized estimator of $\rho$ (dashed-orange) and the \textcite{beirlant2016bias} estimator of $L$ with the penalized estimator of $\rho$ (dashed-blue), have comparable performance. When the tuning parameter $k$ is small, the \textcite{fougeres2015bias} dot aggregated estimator of $L$ with the penalized estimator of $\rho$ (dashed-orange) seems to prevail, while when the tuning parameter $k$ is large, the \textcite{beirlant2016bias} estimator of $L$ with the penalized estimator of $\rho$ (dashed-blue) outperforms under some data generating processes such as Cauchy, Archimax-logistic, and Archimax-mixed.
}

\appendix
\section{Technical Proofs}

\begin{proof}[Proof of Proposition \ref{prop:L_k}]
Let $U_{i}^{(j)}=1-F_{j}(X_{i}^{(j)})$ for $i=1,\dots,n$ and $j=1,\dots,d$. Define, as in \textcite[p. 927]{fougeres2015bias}, 
\[
V_{n,k}(\bx)=\frac{1}{k}\sum_{i=1}^{n}\ind\{U_{i}^{(1)}\leq kx_{1}/n \ \text{or} \ \dots \ \text{or} \ U_{i}^{(d)}\leq kx_{d}/n\}
\]
Further, define, as in \textcite[p. 240]{de2007extreme}, for $\bx \in [0,T]^{d}$,
\[
W_{n}(\bx)=\sqrt{k}\left(V_{n,k}(\bx)-\frac{n}{k}\left\{1-F\left(U_{1}\left(\frac{n}{kx_{1}}\right),\dots, U_{d}\left(\frac{n}{kx_{d}}\right)\right)\right\}\right)
\]
Now let us view $\{W_{n}, n=1,2, \dots\}$ as a random sequence on $\ell^{\infty}([0,T]^{d})$ instead of on the Skorokhod Space. Under Assumption \ref{assump:kspeed},  by \textcite[pp. 240-242]{de2007extreme}, $W_{n}$ as a random sequence on $\ell^{\infty}([0,T]^{d})$ is asymptotically uniformly equicontinuous in probability. If $W_{n}(\bx) = \Op{1}$ for each $\bx\in[0,T]^{d}$, then by \textcite[Theorem 1.5.7]{van1996weak}, $W_{n}$ is asymptotically tight. The remaining proof for Proposition \ref{prop:L_k} follows from the remaining part of \textcite[proof of Proposition 2]{fougeres2015bias}.
 

\end{proof}

\begin{Lemma}\label{lemma:doubleindex}
Suppose that $T>0$, $0<a_{\wedge}\leq a_{\vee}<\infty$, and $Z_{n}\Rightarrow Z$ in $\ell^{\infty}([0,a_{\vee}T]^{d})$. Let $\tZ_{n}$ be a sequence of random functions mapping $[0,T]^{d}\times [a_{\wedge}, a_{\vee}]$ to $\bbR$ defined by $\tZ_{n}(\bx,a)=a^{-1}Z_{n}(a\bx)$ and $\tZ$ be a random function mapping $[0,T]^{d}\times [a_{\wedge}, a_{\vee}]$ to $\bbR$ defined by $\tZ(\bx,a)=a^{-1}Z(a\bx)$. Then $\tZ_{n}\Rightarrow \tZ$ in $\ell^{\infty}([0,T]^{d}\times [a_{\wedge},a_{\vee}])$.
\end{Lemma}
\begin{proof}[Proof of Lemma \ref{lemma:doubleindex}]
Let $\phi: \ell^{\infty}([0,a_{\vee} T]^{d})\to \ell^{\infty}([0,T]^{d}\times [a_{\wedge},a_{\vee}])$ be defined by $\big(\phi(\zeta)\big)(\bx,a)=a^{-1}\zeta(a\bx)$. Now we prove the continuity of $\phi$. Indeed, if 
\[
\sup_{\bx\in [0,a_{\vee}T]^{d}}|\zeta_{n}(\bx)-\zeta(\bx)|\to 0,
\]
then 
\begin{align*}
   \sup_{\bx\in [0,T]^{d}, a\in [a_{\wedge},a_{\vee}]}\Big|\big(\phi(\zeta_{n})\big)(\bx,a)-\big(\phi(\zeta)\big)(\bx,a)\Big|&=\sup_{\bx\in [0,T]^{d}, a\in [a_{\wedge},a_{\vee}]} a^{-1}\Big|\zeta_{n}(a\bx)-\zeta(a\bx)\Big|\\
   &\leq a_{\wedge}^{-1}\sup_{\bx\in [0, a_{\vee}T]^{d}}\Big|\zeta_{n}(\bx)-\zeta(\bx)\Big|\to 0.
\end{align*}
Hence, $\phi$ is continuous. By the Continuous Mapping Theorem, 
\[
\tZ_{n} = \phi(Z_{n})\Rightarrow \phi(Z) = \tZ
\]
in  $\ell^{\infty}([0,T]^{d}\times [a_{\wedge},a_{\vee}])$.
\end{proof}

\begin{proof}[Proof of Proposition \ref{prop:L_ka}]
By \eqref{eq:regv}, Proposition \ref{prop:L_k}, and Lemma \ref{lemma:doubleindex}, for all $T>0$, in $\ell^{\infty}([0,T]^{d}\times [a_{\wedge}, a_{\vee}])$,
\begin{align*}
&\mathph{=}\bigg\{\sqrt{k}\Big(\hL_{ka}(\bx)-L(\bx)-
a^{-1}\alpha(n/k)M(a \bx)
\Big)\bigg\}_{\bx\in [0,T]^{d},a\in [a_{\wedge}, a_{\vee}]}\\
&=\bigg\{\sqrt{k}a^{-1}\Big(\hL_{k}(a\bx)-L(a\bx)-
 \alpha(n/k)M(a \bx)
\Big)\bigg\}_{\bx\in [0,T]^{d},a\in [a_{\wedge}, a_{\vee}]}\Rightarrow \bigg\{a^{-1}Z_{L}(a\bx)\bigg\}_{\bx\in [0,T]^{d}, a\in [a_{\wedge}, a_{\vee}]}.
\end{align*}

\end{proof}

\begin{Lemma}\label{lemma:floored}
Suppose that $T>0$ and $0<a_{\wedge}\leq a_{\vee}<\infty$. Under Assumptions \ref{assump:firorder}, \ref{assump:secorder}, \ref{assump:thiorder}, \ref{assump:negativity}, \ref{assump:smoothness}, and \ref{assump:kspeed},
\[
\sup_{\bx\in [0,T]^{d}, a\in [a_{\wedge},a_{\vee}]}\sqrt{k}\Big|\hL_{ka}(\bx)-\hL_{\floor{ka}}(\bx)\Big|\pto 0.
\]
\end{Lemma}
\begin{proof}[Proof of Lemma \ref{lemma:floored}]
\begin{equation}\label{eq:floored}
    \hL_{ka}(\bx)-\hL_{\floor{ka}}(\bx)= A_{1}+A_{2},
\end{equation}
where
\begin{align*}
    A_{1}&=\frac{1}{\floor{ka}}\sum_{i=1}^{n}\ind_{\big\{X_{i}^{(1)}\geq X_{n-\floor{kax_{1}}+1,n}^{(1)} \ \text{or}\ \dots \ \text{or} \  X_{i}^{(d)}\geq X_{n-\floor{kax_{d}}+1,n}^{(d)}\big\}}\\
    &\mathph{=}-\frac{1}{\floor{ka}}\sum_{i=1}^{n}\ind_{\big\{X_{i}^{(1)}\geq X_{n-\floor{\floor{ka}x_{1}}+1,n}^{(1)} \ \text{or}\ \dots \ \text{or} \  X_{i}^{(d)}\geq X_{n-\floor{\floor{ka}x_{d}}+1,n}^{(d)}\big\}},\\
    A_{2}&=\frac{1}{ka}\sum_{i=1}^{n}\ind_{\big\{X_{i}^{(1)}\geq X_{n-\floor{kax_{1}}+1,n}^{(1)} \ \text{or}\ \dots \ \text{or} \  X_{i}^{(d)}\geq X_{n-\floor{kax_{d}}+1,n}^{(d)}\big\}}\\
    &\mathph{=}-\frac{1}{\floor{ka}}\sum_{i=1}^{n}\ind_{\big\{X_{i}^{(1)}\geq X_{n-\floor{kax_{1}}+1,n}^{(1)} \ \text{or}\ \dots \ \text{or} \  X_{i}^{(d)}\geq X_{n-\floor{kax_{d}}+1,n}^{(d)}\big\}}.
\end{align*}
Since 
\begin{align*}
&\mathph{\subset}\cup_{j=1}^{d}\big\{X_{i}^{(j)}\geq X_{n-\floor{kax_{j}}+1,n}^{(j)}\big\}
\backslash
\cup_{j=1}^{d}\big\{X_{i}^{(j)}\geq X_{n-\floor{\floor{ka}x_{j}}+1,n}^{(j)}\big\}  \\
&\subset
\cup_{j=1}^{d}\Big\{\big\{X_{i}^{(j)}\geq X_{n-\floor{kax_{j}}+1,n}^{(j)}\big\}
\backslash
\big\{X_{i}^{(j)}\geq X_{n-\floor{\floor{ka}x_{j}}+1,n}^{(j)}\big\}\Big\},
\end{align*}
we have that \changeN{uniformly in $\bx\in[0,T]^{d}$ and $a\in [a_{\wedge},a_{\vee}]$},
\begin{equation}\label{eq:floored1}
\begin{aligned}
    |A_{1}|&\leq \frac{1}{\floor{ka}}\sum_{i=1}^{n}\sum_{j=1}^{d}\Big[\ind_{\big\{X_{i}^{(j)}\geq X_{n-\floor{kax_{j}}+1,n}^{(j)}\big\}}-\ind_{\big\{X_{i}^{(j)}\geq X_{n-\floor{\floor{ka}x_{j}}+1,n}^{(j)}\big\}}\Big]\\
    &= \frac{1}{\floor{ka}}\sum_{i=1}^{n}\sum_{j=1}^{d}\ind_{\big\{ X_{n-\floor{kax_{j}}+1,n}^{(j)}\leq X_{i}^{(j)}\leq X_{n-\floor{\floor{ka}x_{j}}+1,n}^{(j)}\big\}}\\
    &\leq \frac{d(T+1)}{\floor{ka}}=\Op{1/k}.
\end{aligned}
\end{equation}
In addition, uniformly in $\bx\in[0,T]^{d}$ and $a\in [a_{\wedge},a_{\vee}]$, by Proposition \ref{prop:L_ka},
\begin{equation}\label{eq:floored2}
    |A_{2}|=\frac{(ka-\floor{ka})}{\floor{ka}}\hL_{ka}=\frac{(ka-\floor{ka})}{\floor{ka}}\Op{1}=\Op{1/k}.
\end{equation}
The lemma follows from \eqref{eq:floored}, \eqref{eq:floored1}, and \eqref{eq:floored2}.
\end{proof}

\begin{Lemma}\label{lemma:M_est}[\textcite{zou2021multiple}] Let $U$ be an arbitrary set and let $(\Theta, d)$ be a metric space. Suppose that $\cL \in \ell^\infty(\Theta)$ is a deterministic function and $(\ccL_{n})_{n\in\bbN}$ is a sequence of random elements in $\ell^\infty(\Theta\times U)$. Assume that $\cL$ has a unique maximizer $\theta_0$ and let
\[
\ctheta_{n}(\bu) \in \argmax_{\theta\in \Theta}\ccL_{n}(\theta;\bu) \quad \forall \, n \in \bbN,\bu \in U;
\]
note in particular that we do not assume that $\ccL_{n}(\cdot;\bu)$ has a unique maximizer. If
\begin{equation}\label{eq:M_cond_1}
\sup_{\bu\in U}\sup_{\theta \in \Theta}\Big|\ccL_{n}(\theta;\bu)-\cL(\theta)\Big|=\op{1},
\end{equation}
then 
\[
\sup_{\bu\in U}\Big[\cL(\theta_{0})-\ccL_{n}(\ctheta_{n}(\bu);\bu)\Big]\leq \op{1}.
\]
If, additionally, for all $\ep>0$, 
\begin{equation}\label{eq:M_cond_2}
\sup_{\theta: d(\theta,\theta_{0})\geq \ep }\cL(\theta)<\cL(\theta_{0}),    
\end{equation}
then
\[
\sup_{\bu\in U}d(\ctheta_{n}(\bu),\theta_{0}) = \op{1}.
\]
\end{Lemma}

\begin{proof}[Proof of Theorem \ref{theo:consist}]
This proof and the proof of Proposition 3.11 in \textcite{zou2021multiple} are similar in spirit \changeN{yet different in technical details}. 
Let
\[
\Big(\tb_{0}(\bx),\tb_{1}(\bx),\trho(\bx)\Big) \in \argmin_{b_{0} , b_{1} \in \bbR, K'\leq r \leq K''} \tRSS(b_{0},b_{1},r;\bx).
\]
Define
\[
\Big(\check b_0(r;\bx),\check b_1(r;\bx)\Big) \coloneqq \argmin_{b_{0}, b_{1}\in\bbR}\tRSS(b_{0},b_{1},r;\bx).
\]
Note that in the minimization problem above $r$ is fixed, whence $\check b_0(r;\bx)$ and $\check b_1(r;\bx)$ can be computed explicitly as the solution of a weighted simple linear regression problem. Since by Assumption \ref{assump:weight}, $\sum_{i \in M_n} w_{n,i} = 1$, standard results (or a tedious computation) show that
\[
\tRSS(\check b_0(r;\bx),\check b_1(r;\bx),r;\bx) = \tilde S_{yy}(\bx)(1 - \tcL_{n}(r;\bx))
\]  
where 
\begin{align*}
\tilde S_{yy}(\bx) 
&\coloneqq 
\sum_{i \in M_n} w_{n,i}\Big\{ \hL_{i}(\bx) - \sum_{j \in M_n} w_{n,j}\hL_{j}(\bx) \Big\}^2,
\\
\tilde S_{xx}(r;\bx) &\coloneqq  \sum_{i \in M_n} w_{n,i}\Big \{ (i/k^{(\rho)})^{-r} - \sum_{j \in M_n} w_{n,j}(j/k^{(\rho)})^{-r} \Big\}^2,
\\
\tilde S_{xy}(r;\bx) 
&\coloneqq 
\sum_{i \in M_n} w_{n,i} \Big\{ \hL_{i}(\bx) - \sum_{j \in M_n} w_{n,j}\hL_{j}(\bx)\Big\} \Big\{ (i/k^{(\rho)})^{-r} - \sum_{j \in M_n} w_{n,j} (j/k^{(\rho)})^{-r} \Big\},
\\
\tcL_{n}(r;\bx) &\coloneqq \frac{\tilde S_{xy}^2(r;\bx)}{\tilde S_{yy}(\bx)\tilde S_{xx}(r;\bx)}.
\end{align*} 
From the definitions above it is clear that
\begin{align*}
& \min_{b_{0}, b_{1} \in \bbR, K'\leq r\leq K''} \hRSS_\eta(b_{0},b_{1},r;\bx)
\\
&= \min_{ K'\leq r\leq K''}\Big( \min_{b_{0}, b_{1} \in \bbR} \tRSS(b_0,b_1,r;\bx) + \frac{\eta}{|r|}\tRSS(\tb_{0}(\bx),\tb_{1}(\bx),\trho(\bx);\bx)\Big) 
\\
&= \tilde S_{yy}(\bx) \min_{ K'\leq r\leq K''}\Big\{ 1 - \tcL_{n}(r;\bx)  + \frac{\eta}{|r|}\Big(1 - \tcL_{n}(\trho(\bx);\bx) \Big)\Big\}. 
\end{align*}
Hence,
\[
\hrho(\bx) \in \argmax_{K'\leq r\leq K''} \Big\{\tcL_{n}(r;\bx) - \frac{\eta}{|r|}\Big(1 - \tcL_{n}(\trho(\bx);\bx) \Big)\Big\}
\]
and by similar but simpler arguments
\[
\trho(\bx) \in \argmin_{K'\leq r \leq K''}\tRSS\Big(\check b_{0}(r;\bx),\check b_{1}(r;\bx),r;\bx\Big) = \argmax_{K'\leq r\leq K''} \tcL_{n}(r;\bx).
\]
\changeN{Next, by Lemma \ref{lemma:floored}, Proposition \ref{prop:L_ka}, and Assumption \ref{assump:weight}, uniformly in $\bx\in \cX$,
\begin{align*}
&\mathph{=}\frac{\sum_{i \in M_n} w_{n,i}\Big\{ \hL_{i}(\bx) -  L(\bx) \Big\}}{\alpha(n/k^{(\rho)})M(\bx)}\\ &= \int_{A} k^{(\rho)}w_{n,\floor{k^{(\rho)}a}}\Bigg\{
\frac{\Bias_{k^{(\rho)},a}(\bx)}{\alpha(n/k^{(\rho)})M(\bx)}+
\frac{\hL_{\floor{k^{(\rho)}a}}(\bx)-L(\bx)-\Bias_{k^{(\rho)},a}(\bx)}{\alpha(n/k^{(\rho)})M(\bx)}
\Bigg\}\diff a\\
&= \int_{A} k^{(\rho)}w_{n,\floor{k^{(\rho)}a}}
\Bigg\{\frac{\Bias_{k^{(\rho)},a}(\bx)}{\alpha(n/k^{(\rho)})M(\bx)}\Bigg\}\diff a +\op{1}\\
&=\int_{A} f(a)a^{-\rho} \diff a +\op{1}.
\end{align*}
By similar and straightforward calculations, }
\begin{align*}
\sup_{\bx \in \cX} \Big|\frac{\tilde S_{yy}(\bx)}{\{\alpha(n/k^{(\rho)})M(\bx)\}^2} - \int_A f(a)(a^{-\rho} - \mu_\rho)^2   \diff a\Big| =\op{1}.
\\
\sup_{K'\leq r\leq K''}\sup_{\bx \in \cX} \Big| \tilde S_{xx}(r;\bx)  - \int_A f(a)(a^{-r} - \mu_r )^2  \diff a\Big| = \op{1}.
\\
\sup_{K'\leq r\leq K''}\sup_{\bx \in \cX} \Big| \frac{\tilde S_{xy}(r;\bx)}{\alpha(n/k^{(\rho)})M(\bx)} - \int_A f(a)(a^{-\rho} - \mu_{\rho})(a^{-r} - \mu_r)   \diff a\Big| = \op{1}.
\end{align*}
where $\mu_r \coloneqq \int_A a^{-r} f(a)  \diff a$. Next, define 
\[
\cL(r) = \frac{\Big\{\int_{A}(a^{-r}-\mu_r)(a^{-\rho}-\mu_{\rho} ) f(a) \diff a \Big\}^{2}}{\int_{A}(a^{-r}-\mu_r)^{2} f(a) \diff a \int_{A}(a^{-\rho}-\mu_{\rho})^{2}f(a) \diff a }.
\]
The arguments given above show that
\begin{equation}\label{eq:unifL}
\sup_{K'\leq r\leq K''}\sup_{\bx \in \cX} |\tcL_{n}(r;\bx) - \cL(r)| = \op{1}.
\end{equation}
Next we show that $\cL$ satisfies~\eqref{eq:M_cond_2} with $\theta_0 = \rho$ and $\Theta = [K',K'']$. Since $\cL$ is continuous, $\{r\in [K',K'']:|r-\rho|\geq \ep\}$ is compact, $\cL(r)\leq 1$ by Cauchy-Schwarz, and $\cL(\rho)=1$, it suffices to show that 
\begin{equation*} 
    \cL(r)=1, \quad \text{only if} \ r=\rho.
\end{equation*} 
This, however, follows again from Cauchy-Schwarz and linear independence of the functions $A \ni a \mapsto a^{-r}$ and $A\ni a \mapsto a^{-\rho}$ for $r \neq \rho$.

Next, apply Lemma \ref{lemma:M_est} with
\begin{equation*}
\Theta=[K',K''], \quad U=\cX,\quad  \ccL_{n}=\tcL_{n},\quad \cL=\cL, \quad\ctheta_{n}=\trho, \quad\theta_{0}=\rho.
\end{equation*}
Condition \eqref{eq:M_cond_1} follows directly from~\eqref{eq:unifL} and hence, by the first part of Lemma \ref{lemma:M_est} and by the fact that $\tcL_{n}(\trho(\bx);\bx) \le 1$ by Cauchy-Schwarz, we have
\[
\tcL_{n}(\trho(\bx);\bx) = \cL(\rho) + \op{1} = 1 + \op{1}
\]
uniformly in $\bx \in \cX$. This implies
\[
\hcL_{n}(r;\bx) \coloneqq \tcL_{n}(r;\bx) - \frac{\eta}{|r|}\Big(1 - \tcL_{n}(\trho(\bx);\bx) \Big) = \cL(r) + \op{1}
\]
uniformly in $r \in [K',K'']$ and $\bx \in \cX$. Hence, we may apply Lemma \ref{lemma:M_est} again, with
\begin{equation*}
\Theta=[K',K''], \quad U=\cX, \quad \ccL_{n}=\hcL_{n}, \quad \cL=\cL, \quad \ctheta_{n}=\hrho^{\mathrm{pen}}, \quad \theta_{0}=\rho,
\end{equation*}
and the result follows by the definition of $\hrho^{\mathrm{pen}}(\bx)$.
\end{proof}

\begin{proof}[Proof of Corollary \ref{coro:consist}]
Since $\cX_{\rho}\subset \cX$, by \eqref{eq:reg_rho} and Theorem \ref{theo:consist},

\[
\left|\hrho_{\cX_{\rho}}^{\mathrm{pen,agg}}-\rho\right| \leq \frac{1}{|\cX_{\rho}|} \sum_{\bx \in \cX_{\rho}} \left|\hrho^{\mathrm{pen}}(\bx)-\rho\right| \leq \sup_{\bx \in \cX_{\rho}}\left|\hrho^{\mathrm{pen}}(\bx)-\rho\right| \leq \sup_{\bx \in \cX}\left|\hrho^{\mathrm{pen}}(\bx)-\rho\right|=\op{1}.
\]

\end{proof}

\begin{proof}[Proof of Proposition \ref{prop:PlugInErrorDot}]
Let $b=(a^{-\rho}+1)^{-1/\rho}$ and $\hb=(a^{-\hrho_{\cX_{\rho}}^{\mathrm{pen,agg}}}+1)^{-1/\hrho_{\cX_{\rho}}^{\mathrm{pen,agg}}}$. By Assumption \ref{assump:negativity}, $\rho<0$. By \eqref{eq:reg_rho}, $\hrho_{\cX_{\rho}}^{\mathrm{pen,agg}}\in [K',K'']\subset(-\infty,0)$. Hence $b>0$ and $\hb>0$. By \eqref{eq:regv} and \eqref{eq:Fou_dot_L},
\[
\sqrt{k}\left(\mL_{k,a}(\bx,\hrho_{\cX_{\rho}}^{\mathrm{pen,agg}})-\mL_{k,a}(\bx,\rho)\right) = \sqrt{k}\left(\hL_{kb}(\bx)-\hL_{k\hb}(\bx)\right) =  B_{1}(\bx) - B_{2}(\bx),
\]
where $B_{1}, B_{2} \in \ell^{\infty}([0,T]^{d})$ are defined by 
\begin{align*}
    B_{1}(\bx)&= \sqrt{k}\left\{ \hL_{kb}(\bx)-L(\bx)-
b^{-1} \alpha(n/k)M(b \bx)
\right\} - \sqrt{k}\left\{ \hL_{k\hb}(\bx)-L(\bx)-
\hb^{-1} \alpha(n/k)M(\hb \bx)
\right\},\\
    B_{2}(\bx)&=\sqrt{k} \alpha(n/k) \left(\hb^{-\rho}-b^{-\rho}\right)M(\bx).
\end{align*}
By Corollary \ref{coro:consist}, we have $\hb\pto b$. Let $a_{\wedge}$ and $a_{\vee}$ be real numbers such that $0<a_{\wedge}< b < a_{\vee}<\infty$. By \textcite[Lemma 1.3.8 (ii) and Theorem 1.5.7]{van1996weak} and Proposition \ref{prop:L_ka}, in probability we have the asymptotically uniform equicontinuity of
\[
\bigg\{\sqrt{k}\Big(\hL_{ka}(\bx)-L(\bx)-a^{-1} \alpha(n/k)M(a \bx)\Big)\bigg\}_{\bx\in [0,T]^{d},a\in [a_{\wedge}, a_{\vee}]}.
\]
Hence, 
\[
B_{1}=\op{1}.
\]
Since $\hb\pto b$, by the Mean Value Theorem, 
\[
|B_{2}|\leq \sqrt{k}\alpha(n/k)(b+\op{1})^{-\rho}|\hb- b| M(\bx)= \Op{\sqrt{k}\alpha(n/k)(\hrho_{\cX_{\rho}}^{\mathrm{pen,agg}}-\rho)}.
\]
\end{proof}

\section*{Acknowledgment}
We thank the authors of \textcite{beirlant2016bias} for sending their codes and thank Axel B\"ucher and Stanislav Volgushev for the fruitful discussion.

\printbibliography
\end{document}